\documentstyle[seceq,epsf,wrapft,preprint]{ptptex}

\newcommand{\idelt}{{\it \Delta}}

\newcommand{\M}[1]{$#1$}

\newcommand{\BM}[1]{\mbox{\boldmath{$#1$}}}

\makeatletter
\def\le{\mathrel{\mathpalette\gl@align<}}
\def\ge{\mathrel{\mathpalette\gl@align>}}
\def\gl@align#1#2{\lower.6ex\vbox{\baselineskip\z@skip\lineskip\z@
    \ialign{$\m@th#1\hfil##\hfil$\crcr#2\crcr=\crcr}}}
\makeatother
\notypesetlogo  
\preprintnumber[3cm]{
January, 1999
}

\title{Scalar-Scalar Ladder Model in the Unequal-Mass Case. III}
\subtitle{Numerical Studies of the \M{P}-Wave Case}

\author{%
Ichio {\sc Fukui}
\footnote{E-mail address: fukui@cc.saga-u.ac.jp}
and
Noriaki {\sc Set\^{o}}$^{**,}$
\footnote{E-mail address: setoh@amath.hiroshima-u.ac.jp}
}

\inst{%
Information Processing Center, Saga University, 
Saga 840-8502, Japan
\\
$^{**}$Department of Applied Mathematics, Faculty of Engineering,
Hiroshima University, Higashi-Hiroshima 739-8527, Japan
}
\recdate{
}

\abst{%
  The eigenvalue problem for the \M{p}-wave bound states formed
 by two unequal-mass scalar particles through the massive scalar 
 particle exchange is analyzed numerically in the framework of 
 the Bethe-Salpeter ladder model.
 As in the \M{s}-wave case, the eigenvalues of the coupling constant
 are found to become complex for some mass configurations in some
 range of the bound state mass. The Bethe-Salpeter amplitudes of 
 the low-lying bound states are also investigated.

}

\begin{document}

\maketitle

\section{Introduction}
The Bethe-Salpeter (BS) formalism is one of the effective (although
now a rather old-fashioned) method to analyze the nonperturbative
aspects in quantum field theory.\cite{rf:Rv2} Usually, in 
the two-body homogeneous BS equation (describing two-particle 
bound states), complete propagators are approximated by free 
propagators and the BS kernel (describing the two-body interaction) 
is approximated by its Born term. 
For the case that the two-body interaction is induced by a particle 
exchange, the approximated BS formalism is referred to as the ladder
model. 
In this model, given a mass of the bound state, the BS equation can 
be interpreted as an eigenvalue equation for the coupling constant.

 Despite of these simplifying approximations, the two-body
homogeneous BS equations cannot be treated analytically except for
the Wick-Cutkosky model.\cite{rf:W_C} This model, describing two
scalar particles (with not neccessarily equal masses) exchanging a
massless scalar particle, can be reduced to an eigenvalue equation of
Sturm-Liouville type. 
It's spectral property is fully clarified analytically as well as 
numerically. 
In the case that the mass of the exchanged particle is nonzero 
(termed as scalar-scalar ladder model by Nakanishi\cite{rf:Rv1}), 
the eigenvalue problem remains, even after the partial wave 
decomposition, a two-dimensional partial differential (or integral)
equation. 
If the two external scalar particles have equal masses, the 
corresponding eigenvalue equation can be reduced to a manifestly 
Hermitian form through the Wick rotation. 
In this case, although the detailed properties of eigenvalues are 
less well understood compared with the Wick-Cutkosky case, 
general properties like the discreteness, reality and positivity
 can be derived from the analysis of the integral kernel.

   Contrary to the equal-mass case above, in the unequal-mass
scalar-scalar ladder model, even the reality of eigenvalues is not 
yet proved or disproved analytically.
The situation was quite confusing nearly
thirty years ago. There were coexisting the papers suggesting 
the reality of 
eigenvalues analytically\cite{rf:N_N} and numerically\cite{rf:Lin},
and also the papers supporting the nonreality of eigenvalues
analytically\cite{rf:Ida} and numerically\cite{rf:Kau}.
Relying on the progress in numerical computational technique,
the present authors performed a rather extensive 
calculation\cite{rf:S_F} to obtain eigenvalues numerically. 
It was found that complex eigenvalues could appear in the first 
and second \M{s}-wave excited states, depending on the mass 
configuration of the two external particles and the exchanged
particle, for the bound state mass squared around the 
pseudothreshold.
The authors also made a numerical analysis of the BS amplitudes
(eigenfunctions)\cite{rf:F_S} to investigate the behavior of 
amplitudes in the momentum space and also to check the numerical 
accuracy of the results obtained in Ref.~\ref{lrf:S_F}.

   The purpose of the present paper is to study the "universality" 
of the complex eigenvalues in the unequal-mass scalar-scalar 
ladder model based on the improved numerical method compared 
with that of Ref.~8. 
As was pointed out in Ref.~8, a preliminary calculation suggested 
that all the eigenvalues of the \M{p}-wave bound states were real, 
for the mass configurations where certain eigenvalues of the 
\M{s}-wave states became complex.  
In the next section, the general formalism for the BS equation 
is presented.  
In \S\,3, the eigenvalue equation in the \M{p}-wave case is 
solved numerically.
It is found that if the mass of exchanged particle is lowered,
complex eigenvalues appear also in the \M{p}-wave case.
In \S\,4, the radial parts of BS amplitudes of some typical bound
states associated with complex as well as real eigenvalues
are depicted in the momentum space.  Physical meaning of the 
results is discussed and a further outlook is made in the final 
section.

\section{General formalism}
 The BS amplitude $\phi(\BM{p},p_4)$ in the momentum space,
with mass squared $s\,(>0)$, formed by two scalar particles with
masses $m_1$ and $m_2$ by exchanging a scalar particle with mass
$\mu$ in the ladder model, obeys, after the Wick rotation,
the following BS equation:
\begin{eqnarray}
\hspace*{-1em}
\lefteqn{\left[m_1^2+\BM{p}^2-(\eta_1\sqrt{s}+ip_4)^2\right]
         \left[m_2^2+\BM{p}^2-(\eta_2\sqrt{s}-ip_4)^2\right]
       \phi(\BM{p},p_4)}
      \makebox[12em] \nonumber \\
&=& \frac{\lambda}{\pi^2}\int d^4p^{\prime}
                       \frac{\phi(\BM{p}^\prime,p_4^\prime)}
    {\mu^2+(\BM{p}- \BM{p}^\prime)^2+(p_4-p^{\prime}_4)^2} \; ,
\label{eq:general}
\end{eqnarray}
\\
with $\eta_1=m_1/(m_1+m_2)$ and $\eta_2=m_2/(m_1+m_2)$.  We shall
denote the model as the $[m_1 \!\Leftarrow\!\mu\!\Rightarrow\! m_2]$ 
model.
 The original setting of the bound state problem was to find the bound
state mass \M{\sqrt{s}} at a given value of the coupling strength
\M{\lambda}. Instead, due to the peculiarity of the ladder model,
we can regard (\ref{eq:general}) as an eigenvalue equation for 
the coupling strength $\lambda$ at a given \M{\sqrt{s}} in the 
range \M{0<\sqrt{s}<m_1+m_2}. 
Taking the unit $m_1+m_2 = 2$ ($\lambda$ has mass dimension +2), 
we can transform Eq.~(\ref{eq:general}), for the 
$[(1+\idelt)\!\Leftarrow\!\mu\!\Rightarrow\!(1-\idelt)]$ model, into
\begin{eqnarray}
\hspace*{-2.5em}
\phi(\BM{p},p_4) & = & \frac{1}{\left[(1\! +\!\idelt)^2
  (1\! -\!\sigma)\! +\! p^2\! -\! 2i(1\! +\!\idelt)
  \sqrt{\sigma}p_4\right]}    \nonumber \\
 & &  \hspace*{-4em}\times
  \frac{1}{\left[(1\! -\!\idelt)^2(1\! -\!\sigma)\!
      +\! p^2\! +\! 2i(1\! -\!\idelt)\sqrt{\sigma}p_4\right]}
            \cdot \frac{\lambda}{\pi^2} \int d^4p^\prime
                   \frac{\phi(\BM{p}^\prime,p_4^\prime)}
                   {\mu^2+(p-p^\prime)^2} \; ,
\label{eq:model}
\end{eqnarray}
with \M{\sigma = s/4} and \M{0 \le \idelt < 1}.
In this unit, the two-particle threshold is equal to 4, while the
pseudothreshold is \M{4\idelt^2}.

The three-dimensional (space part) rotational invariance of
Eq.~(\ref{eq:model}) assures that the four-variable integral
equation can be reduced to a two-variable (e.g. \M{p_4} and \M{p:=
\sqrt{{p_4}^2+\BM{p}^2}}) equation. By using the polar coordinates
in the four-dimensional momentum space \M{(\BM{p},p_4)},
\begin{eqnarray}
\lefteqn{p_4=p\cos\beta,\;\;  \BM{p}=p\sin\beta
        (\sin\theta\cos\varphi,\:
              \sin\theta\sin\varphi,\:\cos\theta)\:,}  \nonumber \\
 & & \hspace{10em}(p \ge 0,\;\; 0 \le \beta,\;\; \theta \le \pi,\;\;
              0 \le \varphi < 2\pi) \nonumber
\end{eqnarray}
through the expansion of the BS amplitude in terms of
four-dimensional spherical harmonics \M{(\sin\beta)^l
C_{L-l}^{l+1}(\cos\beta)Y_{lm}(\theta,\varphi)} with associated
normalization constants \M{N_{L,l}},
\begin{eqnarray}
\phi(\BM{p},p_4) & = & \sum_{L=l}^{\infty}\phi_{L,l}(p)\cdot N_{L,l}
   \cdot(\sin\beta)^l C_{L-l}^{l+1}(\cos\beta)Y_{lm}(\theta,\varphi)
\nonumber
\end{eqnarray}
for suitable \M{l\,(=0,1,2,\ldots)} and \M{m\,(-l \le m \le l)}, 
the two-variable equation is further transformed into an infinite 
system of one-variable integral equations.

By changing the variable \M{p} (the magnitude of the four-dimensional
momentum) to \M{z} defined by
\begin{eqnarray}
p =  \sqrt{\frac{1+z}{1-z}}\:, & \hspace*{2ex} 
                               & z=\frac{p^2-1}{p^2+1}\:,
  \hspace{2em}
  (z:-1 \rightarrow 1 \:\:\mbox{as}\:\: p:0 \rightarrow \infty)
\nonumber
\end{eqnarray}
Eq.~(\ref{eq:model}) is rewritten, after the partial wave
decomposition, as 
\begin{eqnarray}
g_{L,l}(z) & = & \lambda\sum_{L^\prime=l}^{\infty}\int_{-1}^{1}
 dz^\prime\,K_{LL^\prime,l}(z,z^\prime)\;g_{L^\prime,l}(z^\prime).
          \hspace{3em} (L=l,l+1,\ldots)
\label{eq:eigenvalue-eq}
\end{eqnarray}
The \M{l}-th wave BS amplitude \M{\phi(\BM{p},p_4)} is expressed
in terms of \M{g_{L,l}(z)\:}'s as
\begin{eqnarray}
\hspace*{-2.5em}
\phi(\BM{p},p_4) & = & |\BM{p}|^l\sum_{L=l}^{\infty}
(ip)^{L-l}\!\left(\frac{2}{1\! +\! p^2}\right)^{L+3}\!\!
         C_{L-l}^{l+1}\!\left(\frac{p_4}{p}\right)
         g_{L,l}\!\left(\frac{p^2\! -\! 1}{p^2\! +\! 1}\right)
         Y_{lm}(\theta,\varphi).
\label{eq:amplitude}
\end{eqnarray}

The detailed procedure of the above transformations and the
explicit form of the integral kernel matrix functions
\M{K_{LL',l}(z,z')} for \M{l=0} and \M{l=1} are given in
Ref.~\ref{lrf:S_F}.  
They are all expressed as finite sums of elementary functions of 
\M{z} and \M{z'}. 
These matrix elements are all real, but not symmetric 
(\M{K_{L'L,l}(z',z)\neq K_{LL',l}(z,z')}) in general.
If Eq.~(\ref{eq:eigenvalue-eq}) admit a real eigenvalue \M{\lambda},
the corresponding eigenvector \M{g_{L,l}(z)\;\; (L=l,l+1,\ldots)}
can be taken to be real. 
Even in this case, however, since imaginary unit \M{i} appears on 
the right-hand side of Eq.~(\ref{eq:amplitude}),
the BS amplitude \M{\phi(\BM{p},p_4)} generally takes a complex value.
\vspace{1.0em}


\section{Eigenvalues}
In Ref.~\ref{lrf:S_F}, we have analyzed numerically the eigenvalue
problem (\ref{eq:eigenvalue-eq}) for the $s$-wave (\M{l=0}) case.  
We have found there that, for the 
$[1.6 \!\Leftarrow\!1.0\!\Rightarrow\! 0.4]$
model (the mass configuration studied by zur Linden\cite{rf:Lin}
and also by Kaufmann\cite{rf:Kau}) the eigenvalues of the first 
and second excited states form a complex conjugate pair 
in the mass range \M{0.25 < s < 2.65}. 
We shall analyze here the eigenvalue problem for the $p$-wave (\M{l=1}) 
case by a more accurate numerical method than that used in 
Ref.~\ref{lrf:S_F}.  
We have truncated the infinite sum in Eq.~(\ref{eq:eigenvalue-eq}) 
over \M{L^\prime} (from 1 to \M{\infty}) to a finite sum 
(from 1 to a cutoff value \M{L_c}), and replaced the integration over 
\M{z^\prime} by the $N$-point Gauss-Legendre quadrature formula.  
By this approximation, the infinite system of integral equations 
(\ref{eq:eigenvalue-eq}) is reduced to the \M{L_c\cdot N\,}-th order
real but nonsymmetric matrix eigenvalue problem:
\begin{eqnarray}
g_{L,1}(z_j) & = & \lambda\sum_{L^\prime=1}^{L_c}\sum_{k=1}^{N}
    K_{LL^\prime,1}(z_j,z_k)w_k\; g_{L^\prime,1}(z_k)\:. \nonumber \\
             &   & \hspace{5em} (L=1,2,\cdots,L_c,
                   \hspace{1em} j=1,2,\cdots,N)    \nonumber
\end{eqnarray}
The points \M{z_1,\cdots,z_N} are the Gauss-Legendre points on the
interval \M{[-1,1]}, and \M{w_1,\cdots,w_N} are the corresponding
integration weights.

\begin{wrapfigure}{r}{\halftext}
  \epsfysize= 7.2cm
  \epsfbox{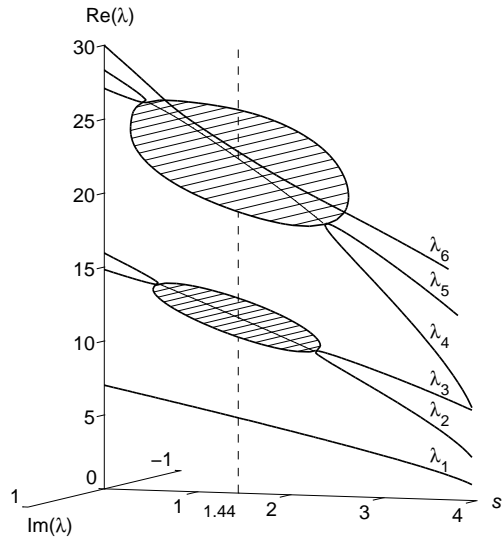}
\caption{The eigenvalues of unequal-mass scalar-scalar model with
         $\Delta = 0.6$ and $\mu = 0.2$. $N$ (the number of
         Gauss-Legendre points) is 47, and $L_c$
         (the four-dimensional angular momentum cutoff) is 10.}
\label{fig:eig}
\end{wrapfigure}

A preliminary numerical calculation made earlier for the $p$-wave
 $[1.6\!\Leftarrow\!1.0\!\Rightarrow\!0.4]$ model suggested that 
 all the eigenvalues were real. To judge whether the emergence of
complex eigenvalues is a peculiar phenomenon to the $s$-wave case,
it is desirable to investigate the eigenvalues in other mass
configurations. 
If we fix $\idelt$ and vary $\mu$, we can expect that complex 
eigenvalues appear more easily for small $\mu$ than for large
$\mu$:  
At $\mu =0$ (the Wick-Cutkosky model), eigenvalues of the first
and second excited states degenerate (forming multiple poles 
consisting of coinciding simple poles in the scattering
Green's function\cite{rf:Nmp}) at $s=4\idelt^2$ (pseudothreshold).
If $\mu$ aquires a small nonzero value, it is probable that the 
degeneracy is removed and two coinciding simple poles become 
a complex conjugate pair for a nonsymmetric real matrix case. 
For this reason we have lowered the value of $\mu$ and investigated 
the eigenvalue  problem for the 
$[1.6\!\Leftarrow\!0.2\!\Rightarrow\!0.4]$ model. 
We have calculated first six eigenvalues, and have found that 
the second and third eigenvalues and also the fourth and fifth 
eigenvalues form  complex conjugate pairs for $s$ in the range 
$0.60<s<2.25$ and  $0.48<s<2.35$, respectively. 
We have sketched in Fig.~\ref{fig:eig} a graph of the first 
six eigenvalues as a function of
$s\,(0<s<4)$ in the complex $\lambda$-plane. 
At $s=4\idelt^2$, for example, \M{\lambda_{2,3} = 11.9 \mp 0.41i} 
and \M{\lambda_{4,5} = 22.6 \mp 0.98i}. 
The ground state \M{\lambda_1} is always real since there is no
partner coinciding with it at $s=4\idelt^2$ and $\mu=0$.  
Also the sixth eigenvalue \M{\lambda_6} remains real: 
The 4-, 5- and 6-th eigenvalues degenerate at $s=4\idelt^2$ 
and $\mu=0$, and the lower two \M{\lambda_4} and \M{\lambda_5} 
form a pair, while \M{\lambda_6} sticks to a life of proud 
loneliness. 
We have also confirmed that in the 
$[1.6\!\Leftarrow\!0.2\!\Rightarrow\!0.4]$ model complex 
eigenvalues certainly appear in the $s$-wave case.

\section{BS amplitudes}
We shall investigate the behavior of the BS amplitude from the 
numerically obtained $p$-wave eigenvectors \M{g^{(k)}_{L,1}(z;s)},
corresponding to the $k$-th eigenvalue \M{\lambda=\lambda_k} 
(\M{k=1, 2, \cdots ,6}) at the bound-state mass squared $s$. 
The radial part (with the factor \M{Y_{lm}(\theta, \phi)} 
omitted) of the approximate BS amplitude 
\M{\phi_k(|\BM{p}|,p_4;s)} is given by
\setcounter{equation}{0}
\begin{eqnarray}
\hspace*{-2.5em}
\phi_k(|\BM{p}|,p_4;s) & = & |\BM{p}|\sum_{L=1}^{L_c}
(ip)^{L-1}\left(\frac{2}{1+p^2}\right)^{L+3}
        C_L^1\!\left(\frac{p_4}{p}\right)
        g^{(k)}_{L,1}\!\left(\frac{p^2-1}{p^2+1};s\right)\:.
\label{eq:rad-amp}
\end{eqnarray}

\begin{figure}[htb]
  \parbox{\halftext}{%
    \epsfysize= 5cm
    \epsfbox{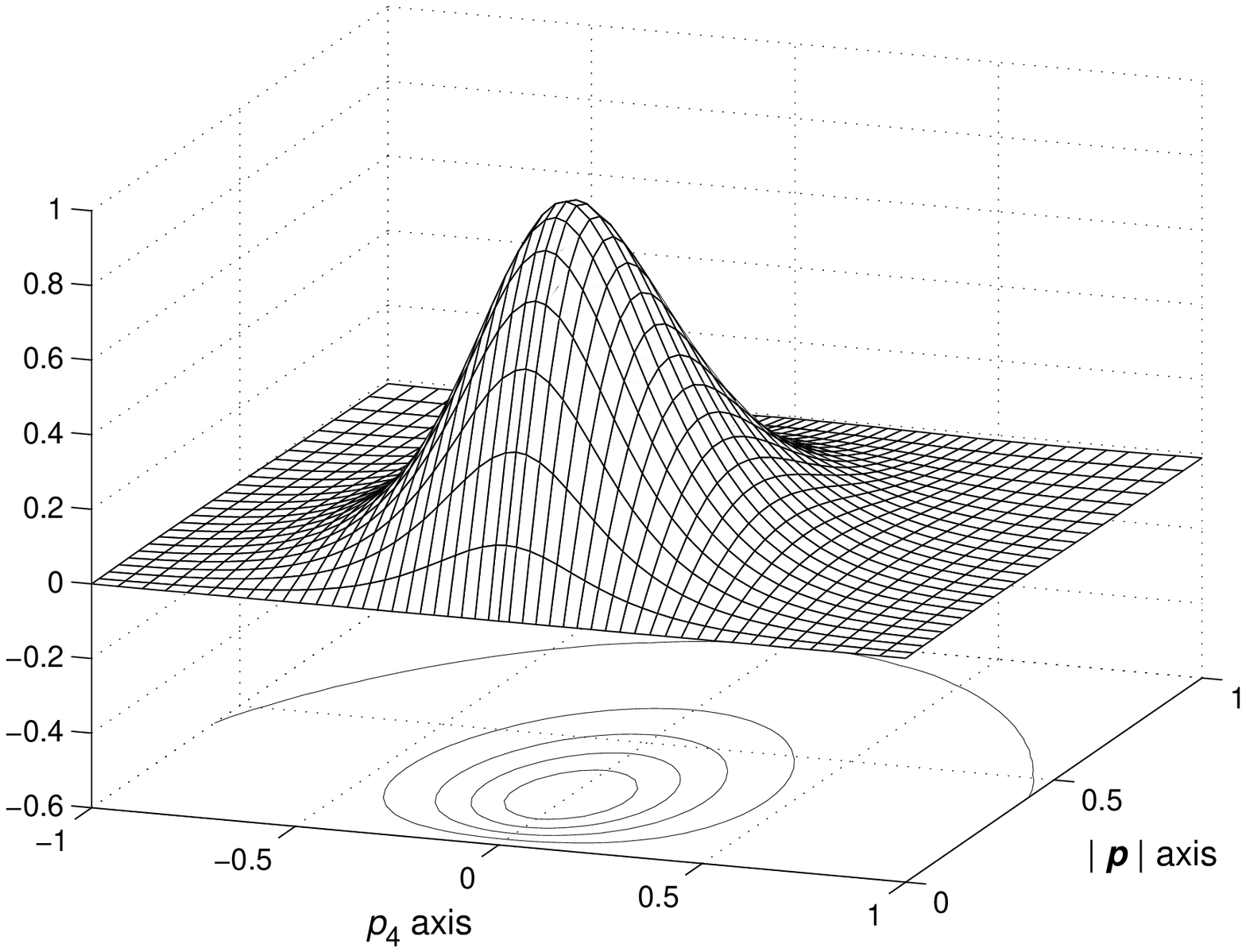}
    \hspace*{8em}{\footnotesize (a)}
    }
  \hspace{8mm}
  \parbox{\halftext}{%
    \epsfysize= 5cm
    \epsfbox{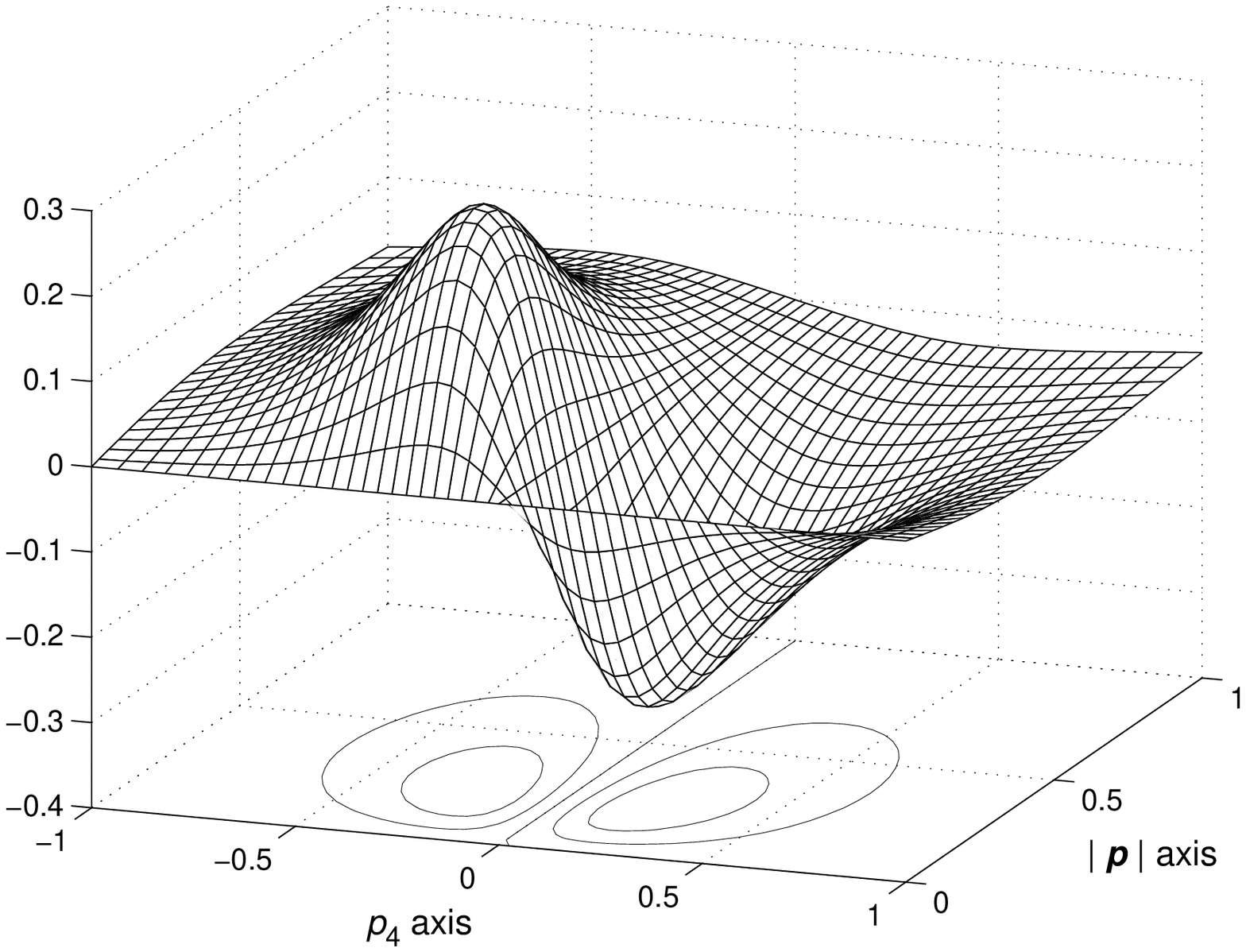}
    \hspace*{8em}{\footnotesize (b)}
    }
  \caption{The BS amplitude of the \M{p}-wave first excited state
    (corresponding to the real eigenvalue) at $s$=0.40,
    (a) real part and (b) imaginary part.}
  \label{fig:wv-real}
\end{figure}

 As the first example, we shall consider a rather deep 
 relativistic case $s=0.40$, where all the eigenvalues are real: 
 \M{\lambda_1 = 6.45}, \M{\lambda_2 = 14.1}, \M{\lambda_3 = 14.7}, 
 \M{\lambda_4 = 26.2}, \M{\lambda_5 = 26.7}, \M{\lambda_6 = 27.7}. 
We can see that $\lambda_2$ and $\lambda_3$ are nearly equal, 
so are the $\lambda_4$, $\lambda_5$ and $\lambda_6$. 
This can be understood as a remnant of the degeneracy between
$\lambda_2$ and $\lambda_3$ (and also among $\lambda_4$, 
$\lambda_5$ and $\lambda_6$ ) at the pseudothreshold in the 
zero-mass exchange case. 
We have plotted in Figs.~\ref{fig:wv-real}(a) and 
\ref{fig:wv-real}(b) the real and 
imaginary parts of the BS amplitude of the first excited state 
($k = 2$) on the \M{p_4}-\M{|\BM{p}|}

\noindent  half plane. The complex-valued 
amplitude is normalized so that 
at the point where the absolute value becomes largest, the real 
part of the magnitude is equal to 1.00, 
while its imaginary part is zero.
We can see that due to the presence of the kinematical factor 
\M{|\BM{p}|} in Eq.~(\ref{eq:rad-amp}), the BS amplitude 
vanishes on the \M{p_4}-axis and peak of the amplitude is pushed 
into the inner \M{|\BM{p}|} region. 
The real (imaginary) part is an even (odd) function of 
the variable \M{p_4}.

\begin{figure}[htb]
  \parbox{\halftext}{%
    \epsfysize= 5cm
    \epsfbox{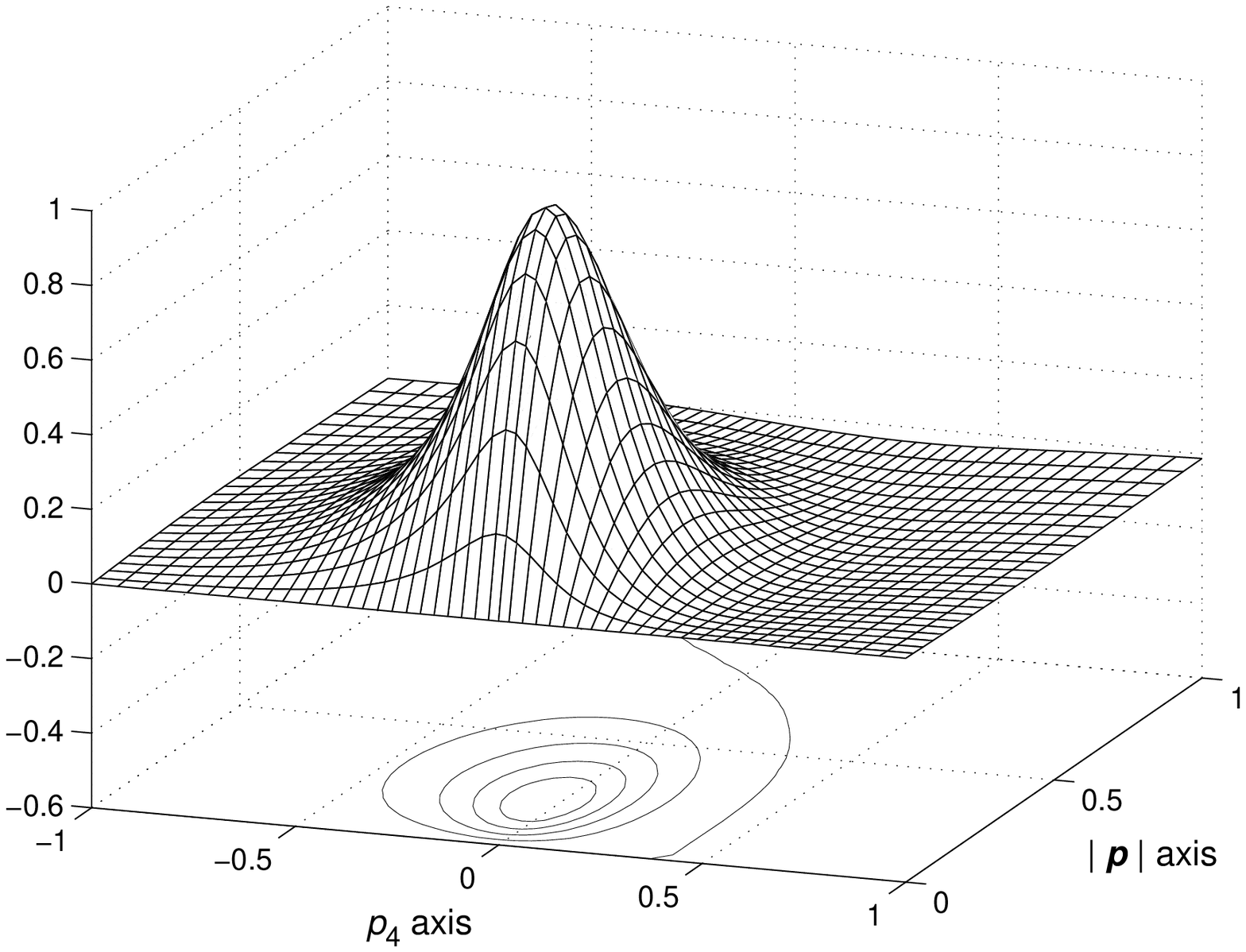}
    \hspace*{8em}{\footnotesize (a)}
    }
  \hspace{8mm}
  \parbox{\halftext}{%
    \epsfysize= 5cm
    \epsfbox{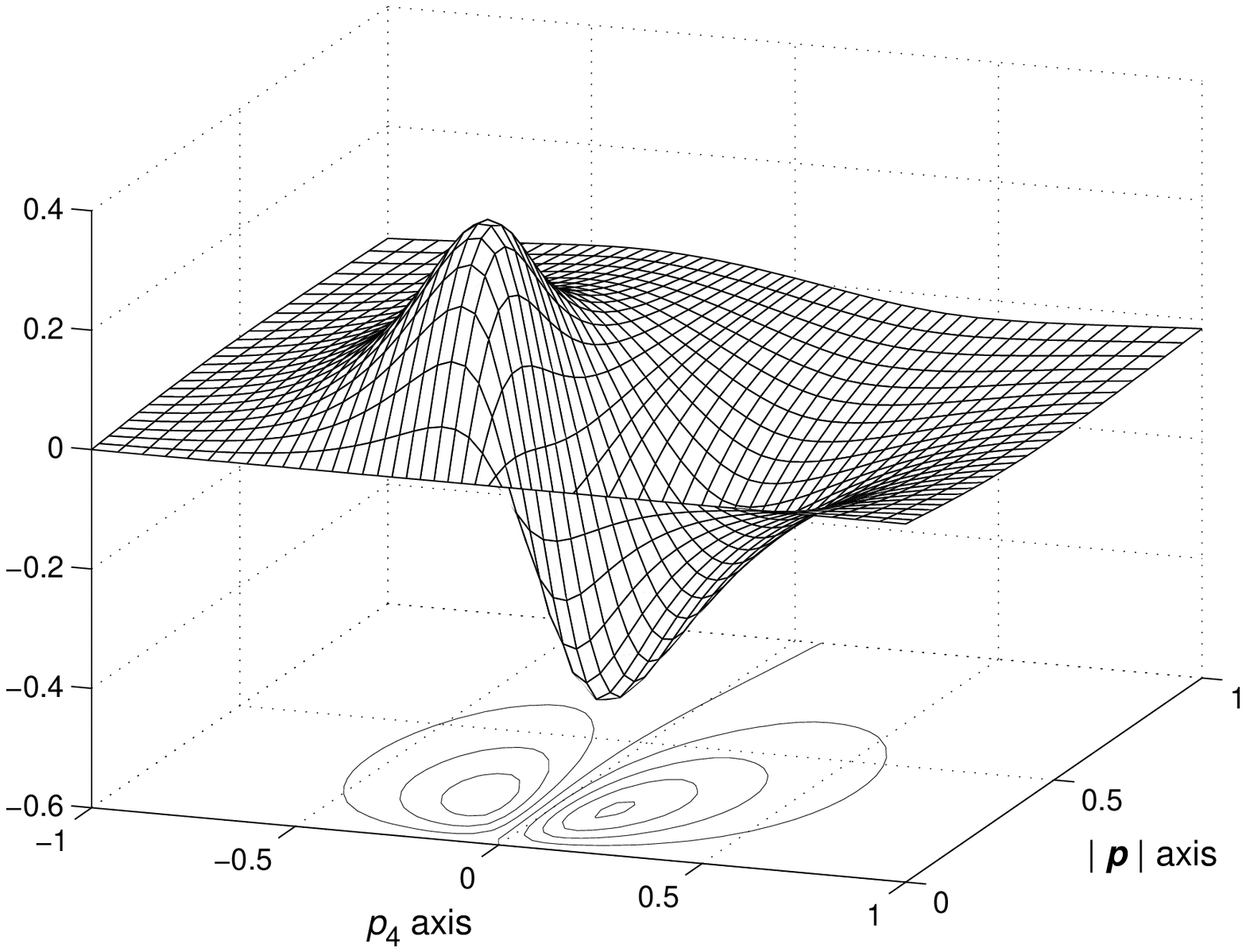}
    \hspace*{8em}{\footnotesize (b)}
    }
  \caption{The BS amplitude of the \M{p}-wave first excited state
    (corresponding to the complex eigenvalue) at $s$=1.44,
    (a) real part and (b) imaginary part.}
  \label{fig:wv-imag}
\end{figure}

The real and imaginary parts of the BS amplitude corresponding the
complex eigenvalue \M{\lambda_2= 11.9 - 0.41i} at \M{s=1.44}
(pseudothreshold) are depicted in Figs.~\ref{fig:wv-imag}(a) and 
\ref{fig:wv-imag}(b), respectively.
Other eigenvalues are \M{\lambda_1 = 5.1}, 
\M{\lambda_{4,5} = 22.6 \mp 0.98i} and \M{\lambda_6 = 23.1}.
The contour plot on the horizontal plane shows that the symmetry 
(antisymmetry) of the real
(imaginary) part of the amplitude under the \M{p_4}-inversion 
is broken.

\begin{figure}[htb]
  \parbox{\halftext}{%
    \epsfysize= 5cm
    \epsfbox{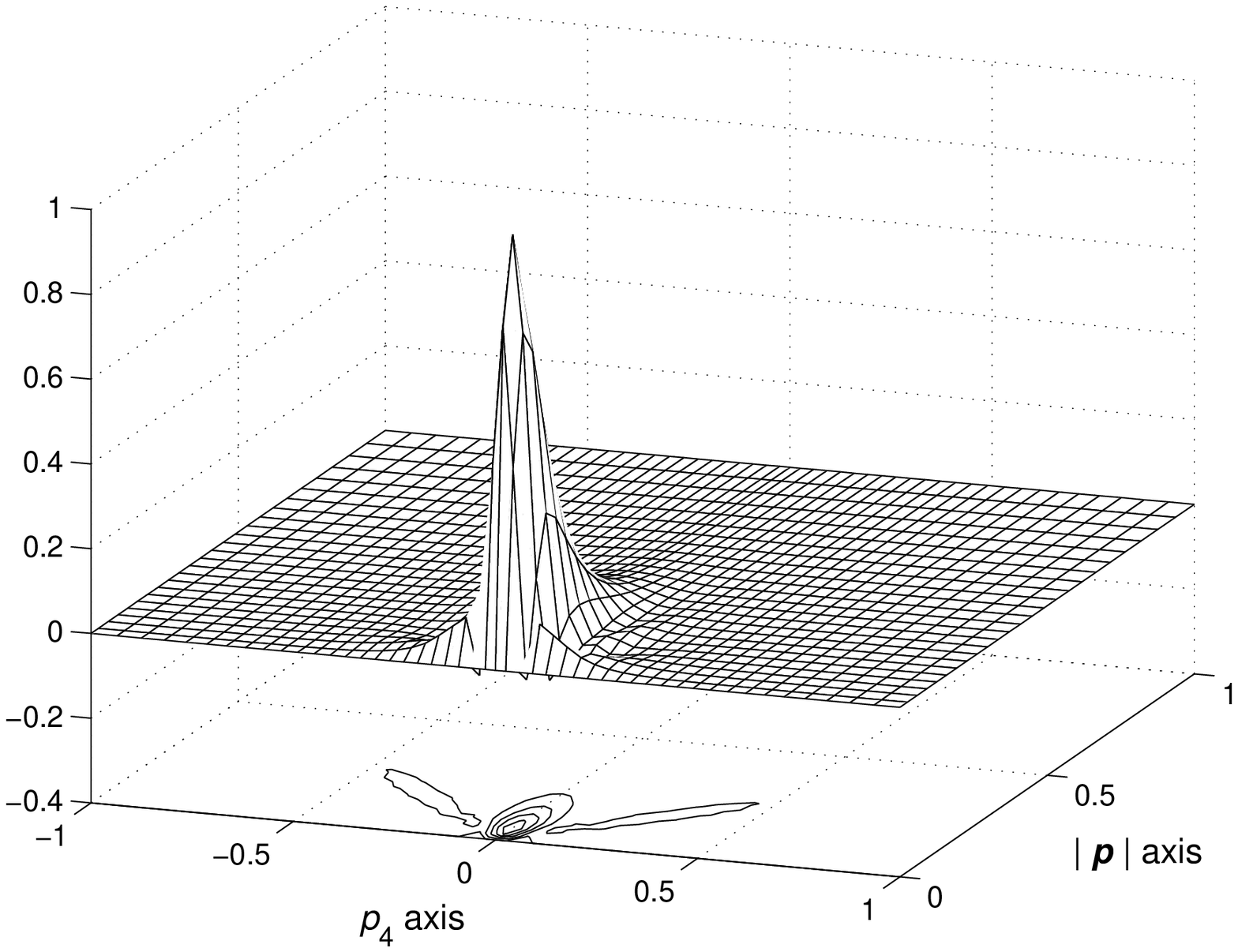}
    \hspace*{8em}{\footnotesize (a)}
    }
  \hspace{8mm}
  \parbox{\halftext}{%
    \epsfysize= 5cm
    \epsfbox{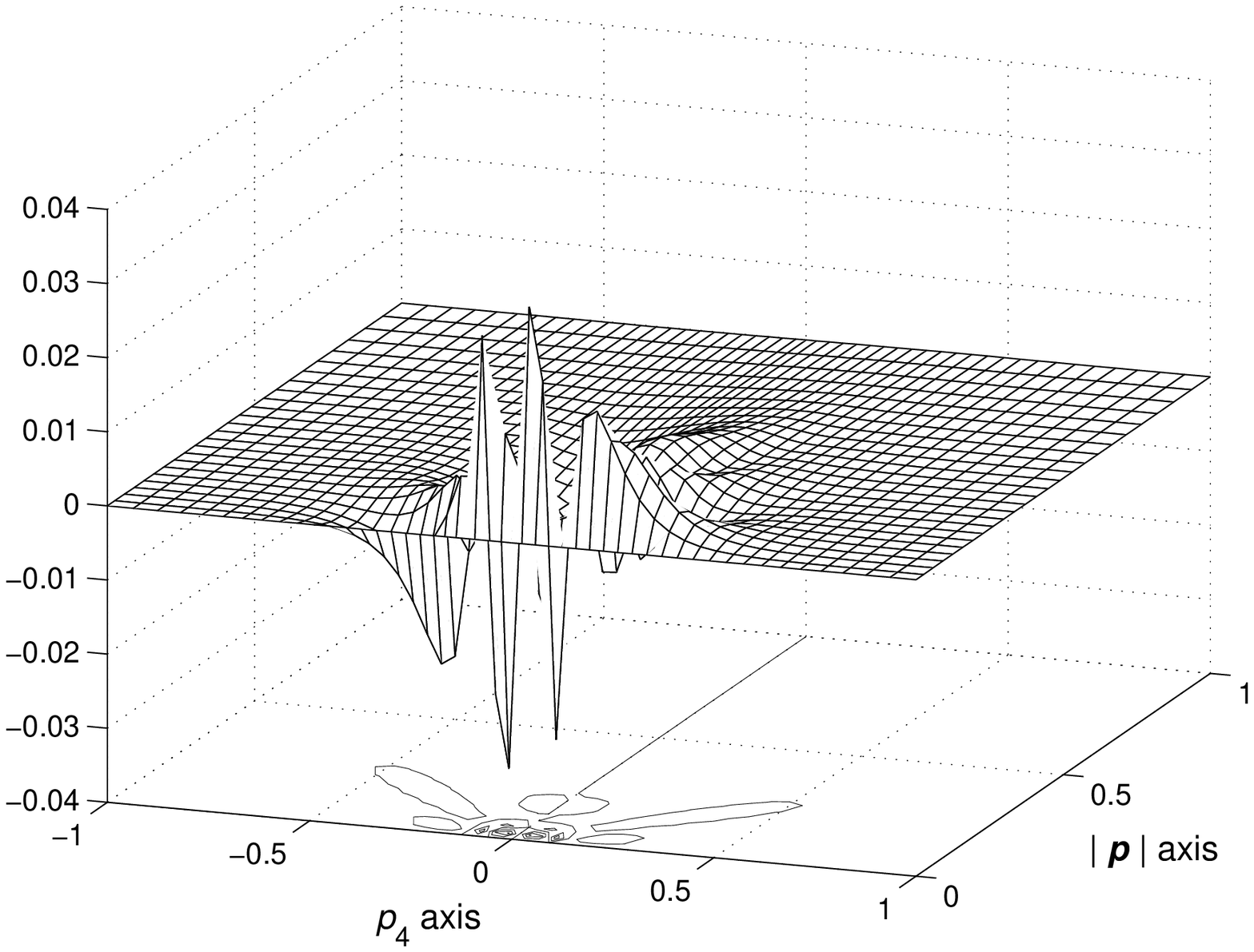}
    \hspace*{8em}{\footnotesize (b)}
    }
  \caption{The BS amplitude of the \M{p}-wave ground state
    at $s$=3.90, (a) real part and (b) imaginary part.
    Note that the scale of the vertical axis in
    Fig.~\ref{fig:wv-nonr}(b) is compressed to one tenth compared
    with that in Fig.~\ref{fig:wv-imag}(b).}
  \label{fig:wv-nonr}
\end{figure}

As the last example, the BS amplitude of the ground state at 
the nearly nonrelativistic energy \M{s=3.90} is plotted in 
Fig.~\ref{fig:wv-nonr}.
The corresponding eigenvalue is \M{\lambda_1 = 1.2}.
The eigenvalue

\noindent  approaches zero in the nonrelativistic limit 
(\M{s \uparrow 4}), since the ground state have a well-defined 
limit (normal state). At this energy, the \M{p_4}-symmetric real part dominates over 
the \M{p_4}-antisymmetric imaginary part: 
The maximum of the imaginary part is at most several percent of 
the maximum of the real part. 
Moreover, if we integrate the amplitude over the \M{p_4} variable, 
which corresponds to the equal-time nonrelativistic wave function, 
the contribution from the imaginary part vanishes identically. 
As in the \M{s}-wave case, the support of the amplitude is 
concentrated near the origin.

\section{Summary and discussion}
  We have thus confirmed that the emergence of complex 
eigenvalues (of the coupling constant) is a commonplace event in 
the scalar-scalar ladder model in the unequal-mass case.  
For a fixed mass difference \M{2\idelt} between two constituent 
paticles, complex eigenvalues in the \M{p}-wave case appear in 
the narrower mass range of the exchanged particle than that in 
the \M{s}-wave case. 
It is expected that complex eigenvalues also appear in higher waves 
if the exchanged-particle mass \M{\mu} is nonzero and sufficiently 
small.  
This is due to the degeneracy of eigenvalues of the Wick-Cutkosky 
model at the pseudothreshold \M{s=4\idelt^2}: 
A pair of eigenvalues, which is degenerate at \M{s=4\idelt^2}
when \M{\mu =0}, is attracted into a complex conjugate pair 
when \M{\mu} becomes nonzero, 
because of the non-Hermiticity of the matrix equation
when \M{\idelt} is not equal to zero.
  Based on these considerations, it will be possible to prove 
analytically (or at least semi-analytically) the existence of 
complex eigenvalues at \M{s=4\idelt^2}, with the aid of 
perturbation theory around \M{\mu =0}.





\end{document}